\documentclass[10pt,conference]{IEEEtran}

\usepackage{epsfig,setspace,amsmath,epsf,amssymb,bm,theorem,cite, graphicx, epstopdf, algorithm, algpseudocode}

\newtheorem{theorem}{Theorem}
\newtheorem{example}{Example}

\newenvironment{Proof}[1]{\medskip\par\noindent{\bf Proof:\,}\,#1}{{\mbox{\,$\blacksquare$}\par}}
\newcommand*{\QEDA}{\hfill\ensuremath{\blacksquare}}

\IEEEoverridecommandlockouts
\allowdisplaybreaks

\begin{document}

\title{Novel Decentralized Coded Caching through \\ Coded Prefetching\thanks{This work was supported by NSF Grants CNS 13-14733, CCF 14-22111, and CNS 15-26608.}}

\author{\IEEEauthorblockN{Yi-Peng Wei \qquad Sennur Ulukus}
\IEEEauthorblockA{Department of Electrical and Computer Engineering\\
University of Maryland College Park, MD 20742\\
{\it ypwei@umd.edu \qquad ulukus@umd.edu} }}

\maketitle

\begin{abstract}
We propose a new \emph{decentralized} coded caching scheme for a two-phase caching network, where the data placed in user caches in the prefetching phase are random portions of a maximal distance separable (MDS) coded version of the original files. The proposed scheme achieves a better rate memory trade-off by utilizing the reconstruction property of MDS codes which reduces the number of transmissions that are useful only for a small subset of users in the delivery phase. Unlike the previously available coded prefetching schemes, the proposed scheme does not require to have more users than files. The proposed scheme can be viewed as a generalization of the original uncoded prefetching based decentralized coded caching scheme, and likewise, is applicable to various network topologies.
\end{abstract}

\section{Introduction}

Consider a two-phase caching network consisting of one server with $N$ files connecting to $K'$ users through an error-free shared link \cite{maddah2014fundamental, maddah2015decentralized}. Each user has a local cache memory which can store $M$ files. The two phases are the placement phase and the delivery phase. In the placement phase, when the network traffic load is low, each user can access the entire $N$ files in the server and fill their cache memories in advance. In the delivery phase, when the network traffic load is high, random $K$ out of $K'$ users request a file from the server individually. The server delivers messages through the error-free shared link to the $K$ users. The request of each user is unknown a priori in the placement phase. Each user reconstructs the file they requested by using the messages sent from the server and the side information stored in their cache memory. The objective is to minimize the traffic load in the delivery phase due to the high traffic load then.

The two-phase caching network is first studied in \cite{maddah2014fundamental}, where by a proper assignment of the local cache memory content in the placement phase, coded-multicasting opportunities are created in the delivery phase, reducing the delivery rate. In \cite{maddah2014fundamental}, the number of users, $K'$, in the placement phase is equal to the number of users, $K$, in the delivery phase, and therefore, the server can carefully coordinate each user's local cache memory. This is referred to as \textit{centralized coded caching}. In \cite{maddah2015decentralized}, a more general scenario is considered where the number of users, $K'$, in the placement phase is not the same as the number of users, $K$, in the delivery phase. Since an arbitrary subset of $K'$ users can make a request in the delivery phase and this number $K$ is unknown a priori in the placement phase, the server cannot carefully coordinate each user's cache content as in \cite{maddah2014fundamental}. In \cite{maddah2015decentralized}, an independent and identical random caching scheme is proposed, and is referred to as \textit{decentralized coded caching}. Through the independent and identical random caching scheme in \cite{maddah2015decentralized}, coded-multicasting opportunities are still created in the delivery phase. Decentralized coded caching is widely applicable in other contexts, such as, online coded caching \cite{pedarsani2016online}, where the cached content of users is updated during the delivery phase; and various  network topologies such as hierarchical \cite{karamchandani2016hierarchical} and general \cite{naderializadeh2017optimality} networks.

Both centralized coded caching in \cite{maddah2014fundamental} and decentralized coded caching in \cite{maddah2015decentralized} are uncoded prefetching schemes \cite{wan2016optimality,yu2016exact}, i.e., each user stores a subset of the bits of the original files. For centralized uncoded prefetching, with $N \geq K$,  \cite{wan2016optimality} determines the exact rate memory trade-off for the worst-case file requests. For both centralized and decentralized uncoded prefetching, \cite{yu2016exact} determines the exact rate memory trade-off for arbitrary $N$, $K$ and $M$ both for the worst-case and average file requests. For coded prefetching, the state of the art order optimality result is of factor $2$ as presented in \cite{yu2017characterizing}.

For centralized setting, reference \cite[Appendix]{maddah2014fundamental} provides an example to show that coded prefetching outperforms uncoded prefetching. For the case of more users than files, i.e., $N \leq K$, better coded caching schemes are provided in \cite{chen2014fundamental, tian2016caching, gomez2016fundamental}. The coded caching scheme in \cite{chen2014fundamental} can only be applied to $M\leq \frac{1}{K}$. The achievable rate memory pair in \cite{chen2014fundamental} is also achieved by a more general coded prefetching scheme given in \cite{tian2016caching}, which is based on rank metric codes and MDS codes. In \cite{gomez2016fundamental}, a novel coded caching scheme is provided for $\frac{1}{K} \leq M\leq \frac{N}{K}$. More detailed comparisons can be found in \cite{tian2017uncoded}.

In this work, we propose a novel decentralized coded caching scheme based on coded prefetching\footnote{The coded prefetching schemes in \cite[Appendix]{maddah2014fundamental} and \cite{chen2014fundamental, tian2016caching, gomez2016fundamental} perform coding over different files, i.e., what is prefetched is a mix of all files, while our coded prefetching here performs coding over the symbols in each file, i.e., what is prefetched is a mix of all symbols in individual files.}. The proposed scheme first MDS codes each file in the server. Then, each user independently and randomly caches the MDS coded file as in \cite{maddah2015decentralized}. By utilizing the reconstruction property of MDS codes, the proposed scheme reduces the transmission of the messages beneficial only to a small subset of users in the delivery phase. This yields an improved rate memory trade-off curve. Different from the existing coded prefetching schemes for centralized setting \cite{chen2014fundamental, tian2016caching, gomez2016fundamental}, the proposed scheme works for arbitrary $N$ and $K$. In fact, for the decentralized setting, one cannot assume $N \leq K$, as the number of users making requests in the delivery phase is uncertain. The proposed scheme can be viewed as a generalization of the original uncoded prefetching decentralized coded caching scheme in \cite{maddah2015decentralized}, and is likewise, applicable to online coded caching \cite{pedarsani2016online} and general network topologies \cite{karamchandani2016hierarchical, naderializadeh2017optimality}.

\section{System Model and Background} \label{Sec_system}

\begin{figure}
\centering
\epsfig{file=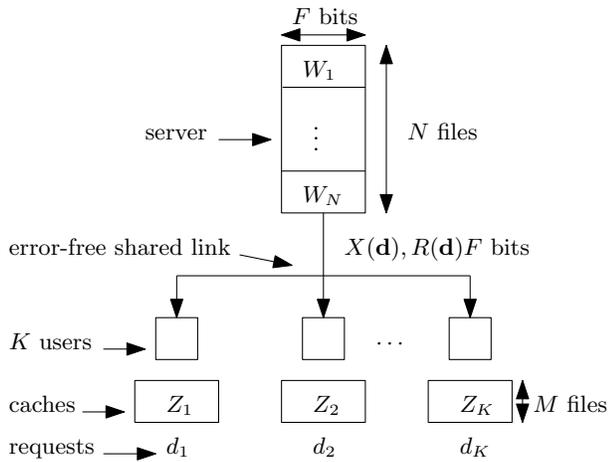,width=0.44\textwidth}
\caption{Caching network.}
\label{Fig_system}
\vspace*{-0.5cm}
\end{figure}

We consider a caching network consisting of one server and $K'$ users. The server is connected to the $K'$ users through an error-free shared link. The server has $N$ files denoted by $W_1, W_2, \dots, W_N$. Each file is of size $F$ bits. Each user has a local cache memory $Z_k$ of size $MF$ bits for some real number $M\in[0,N]$. There are two phases in this network, a placement phase and a delivery phase. In the placement phase, the traffic load is low. User $k$ can access all the $N$ files and fill its cache memory $Z_k$. Therefore, $Z_k=\phi_k(W_1, W_2, \dots, W_N)$, where $\phi_k: \mathbb{F}_2^{NF} \rightarrow \mathbb{F}_2^{MF}$. In the delivery phase, the traffic load is high. Random $K$ users out of total $K'$ users request a file individually from the server (see Fig.~\ref{Fig_system}). Let us denote each user's request as $d_k$, and the request vector as $\mathbf{d}=(d_1, \dots, d_K)$. The server outputs $X(\mathbf{d})$ of size $R(\mathbf{d})F$ bits through the error-free shared link to the $K$ users, where $R(\mathbf{d})$ refers to the load of the network in the delivery phase. A rate is said to be achievable if each user $k$ can decode the file $W_{d_k}$ by utilizing $X(\mathbf{d})$ and $Z_k$. In this work, we study the worst-case normalized rate defined as
\begin{align}
R(M,N,K)\triangleq \max_{d_1, \dots, d_K} R(\mathbf{d}).
\end{align}
The objective is to minimize the rate in the delivery phase.

In the placement phase, which files will be requested by the users in the delivery phase is unknown in advance, i.e., the cache memory $Z_k$ is determined before knowing $\mathbf{d}$. Moreover, which $K$ users out of the $K'$ users will make a request is also unknown a priori. Therefore, in the placement phase, the server fills the local cache memory of each user through a symmetric and random process. Next, we use an example to illustrate the decentralized coded caching algorithm in \cite{maddah2015decentralized}.

\begin{example} \label{ex1}
\normalfont Consider a two-phase caching network with $N=2$, $M=1$ and $K'$ users. We denote the two files as $A$ and $B$. In the placement phase, each user has a space of $\frac{MF}{N}=\frac{F}{2}$ bits in its cache to devote for each file. Therefore, each user independently and randomly chooses $\frac{F}{2}$ bits of each file to cache, and the memory size constraint is satisfied.

In the delivery phase, we consider the case $K=2$ with $d_1=A$ and $d_2=B$. We partition each file as
\begin{align} \label{set_partition}
A=(A_\emptyset, A_1, A_2, A_{1,2}),
\end{align}
where $A_S$ denotes the bits of file $A$ that are cached at the users in the set $S$. To satisfy the request of each user, the server first sends out $A_2 \oplus B_1$. Since $B_1$ is cached by the first user, the first users can obtain $A_2$. Similarly, the second user obtains $B_1$. Thus, a coded-multicasting opportunity is obtained even though the setting is decentralized. Next, the server sends out $A_\emptyset$ and $B_\emptyset$. Combined with the local cache memory $(A_1, A_{1,2})$, the first user gets file $A$. Similarly, the second user gets file $B$.

The traffic load of the delivery phase is calculated using the law of the large numbers. For the file size $F$ large enough,
\begin{align}
&|A_\emptyset| \approx F \left(1-\frac{1}{2}\right)\left(1-\frac{1}{2}\right)=\frac{F}{4}, \\
&|A_1|         \approx F \left(\frac{1}{2}\right)\left(1-\frac{1}{2}\right)=\frac{F}{4},   \\
&|A_2|         \approx F \left(1-\frac{1}{2}\right) \left(\frac{1}{2}\right)=\frac{F}{4}, \\
&|A_{1,2}|     \approx F \left(\frac{1}{2}\right) \left(\frac{1}{2}\right) =\frac{F}{4}.
\end{align}
Therefore, the total number of transmitted bits is
\begin{align}
|A_2 \oplus B_1|+|A_\emptyset|+|B_\emptyset|=\frac{3}{4}F,
\end{align}
and the normalized traffic load is $\frac{3}{4}$. By inspecting all possible file requests, we note that the worst-case normalized rate is $\frac{3}{4}$ for the decentralized coded-caching algorithm in \cite{maddah2015decentralized}. \QEDA
\end{example}

For uncoded prefetching, the exact rate memory trade-off for the decentralized setting is \cite{yu2016exact} 
\begin{align} \label{exact_un}
R_{\mathrm{u,dec}}(M,N,K)=\frac{N-M}{M} \left( 1- \left(\frac{N-M}{N}\right)^{\min\{N,K\}} \right).
\end{align}
For uncoded prefetching, the exact rate memory trade-off for the centralized setting \cite[Corollary 1]{yu2016exact} is
\begin{align} \label{exact_cen_un}
R_{\mathrm{u,cen}}(M,N,K)=\frac{\binom{K}{r+1}-\binom{K-\min\{N,K\}}{r+1} }{\binom{K}{r}},
\end{align}
where $r=\frac{KM}{N}$, and $r\in\{0,1,\dots, K\}$. When $r \notin \mathbb{Z}^+\cup 0$, $R_{\mathrm{u,cen}}(M,N,K)$ is equal to the lower convex envelope \cite{yu2016exact}.

\section{Novel Decentralized Coded Caching}

We start with a motivating example to illustrate the main idea of the proposed new decentralized coded caching scheme.

\begin{example}
\normalfont Consider the same setting as in Example~\ref{ex1} with $N=2$, $M=1$, and $K=2$. Note that in the delivery phase of Example~\ref{ex1}, although $A_2 \oplus B_1$ is simultaneously useful for the two users, $A_\emptyset$ and $B_\emptyset$ are not. In this example, we try to lower the traffic load by reducing the messages useful only  for one of the users through the assistance of MDS codes.

Consider a $(2F, F)$ MDS code over\footnote{To apply $(2F, F)$ MDS over $\mathbb{F}_q$, we consider the file size to be $F \log_2 q$ bits. Since $M=1$, the cache size is of $F \log_2 q $ bits. When we compare with Example~\ref{ex1}, we also consider the file size to be $F \log_2 q $ in Example~\ref{ex1}.} $\mathbb{F}_q$. We transform each file with $F \log_2 q $ bits into $F$ symbols over $\mathbb{F}_q$. Then, we encode each transformed file into an MDS coded file with size $2F$ symbols. From the MDS property, arbitrary $F$ symbols reconstruct the whole file. In the placement phase, each user independently and randomly chooses $\frac{MF}{N}=\frac{F}{2}$ symbols of each coded file to cache, and the memory size constraint is satisfied. Here, we denote the MDS coded files as $A'$ and $B'$. As in Example~\ref{ex1}, we partition each coded file as
\begin{align} \label{set_partition_prime}
A'=(A'_\emptyset, A'_1, A'_2, A'_{1,2}).
\end{align}
By law of large numbers, for $F$ large enough, we have
\begin{align}
&|A'_\emptyset| \approx 2F \left(1-\frac{1}{4}\right)\left(1-\frac{1}{4}\right)=\frac{9}{8}F, \\
&|A'_1|         \approx 2F \left(\frac{1}{4}\right)\left(1-\frac{1}{4}\right)=\frac{3}{8}F,   \\
&|A'_2|         \approx 2F \left(1-\frac{1}{4}\right) \left(\frac{1}{4}\right)=\frac{3}{8}F, \\
&|A'_{1,2}|     \approx 2F \left(\frac{1}{4}\right) \left(\frac{1}{4}\right) =\frac{1}{8}F.
\end{align}

In the delivery phase, we again consider the case $d_1=A$ and $d_2=B$. The server first sends out $A'_2 \oplus B'_1$. Since the first user has $B'_1$ in the local cache, the first user can get $A'_2$. Similarly, the second user gets $B'_1$. Since $|A'_2|=|B'_1|=\frac{3}{8}F$, each user till now has $\frac{7}{8}F$ symbols of the file they requested. Therefore, the server only needs to send out $\frac{1}{8}F$ symbols of $|A'_\emptyset|$ and  $\frac{1}{8}F$ symbols of $|B'_\emptyset|$. In total, the normalized traffic load is $\frac{5}{8}$, which is better than  $\frac{3}{4}$ in Example~\ref{ex1}. 

We can further boost the performance by using a $(3F,F)$ MDS code or a $(4F, F)$ MDS code. Through a similar calculation, we can show that the normalized traffic loads are $\frac{7}{12}$ and $\frac{9}{16}$, respectively, in these cases. For a general $(rF, F)$ MDS code, the normalized traffic load is $\frac{1}{2}+\frac{1}{4r}$. As $r\rightarrow \infty$, $R\rightarrow\frac{1}{2}$, which is the normalized traffic load for centralized coded caching given in \eqref{exact_cen_un}. We see in this example that, for this case, the performance of this system improves monotonically in $r$, and as $r\rightarrow \infty$ the performance of the decentralized system reaches the performance of the centralized system. \QEDA
\end{example}

For some cases, the server does not need to send out $|A'_\emptyset|$.

\begin{example} \label{ex3}
\normalfont Consider the two-phase caching network with $N=2$, $M=1$ and $K=3$. Consider a $(2F, F)$ MDS code over $\mathbb{F}_q$. Encode each file using this MDS into a file of size $2F$. In the placement phase, each user independently and randomly chooses $\frac{F}{2}$ symbols of each coded file to cache, and the memory size constraint is satisfied. 

We partition each coded file as
\begin{align} \label{set_partition_EX3}
A'=(A'_\emptyset, A'_1, A'_2, A'_3, A'_{1,2}, A'_{2,3}, A'_{1,3}, A'_{1,2,3}).
\end{align}
For $F$ large enough, each subfile's size is about
\begin{align}
&|A'_{1,2,3}| \approx 2F\left(\frac{1}{4} \right)\left(\frac{1}{4}\right)\left(\frac{1}{4}\right)=\frac{1}{32}F, \\
&|A'_{1,2}|\approx|A'_{2,3}|\approx|A'_{1,3}| \approx 2F\left(\frac{1}{4} \right)\left(\frac{1}{4} \right) \left(\frac{3}{4} \right)=\frac{3}{32}F, \\
&|A'_1|\approx|A'_2|\approx|A'_3| \approx 2F \left(\frac{1}{4} \right) \left(\frac{3}{4} \right)\left(\frac{3}{4} \right) = \frac{9}{32}F, \\
&|A'_\emptyset| \approx 2F \left(\frac{3}{4} \right) \left(\frac{3}{4} \right) \left(\frac{3}{4} \right) = \frac{27}{32}F.
\end{align}

In the delivery phase, we consider the case $d_1=A$, $d_2=B$ and $d_3=A$. The server first sends out $A'_{2,3} \oplus B'_{1,3} \oplus A'_{1,2}$. Since the first user has $(B'_{1,3},A'_{1,2})$ in the local cache, $A'_{2,3}$ can be decoded correctly. Similar argument holds for the other two users. Next, the server sends out $A_2' \oplus B_1'$, $A_2' \oplus B_3'$ and $A_3' \oplus A_1'$. Since the first user has $(B_1', A_1')$ in the local cache, $A_2'$ and $A_3'$ can be obtained. Till now, the first user has every subfile of $A'$ except $|A'_\emptyset|$. Therefore, the first user has $\frac{37}{32}F$ symbols of coded file $A'$. This is sufficient to reconstruct the requested file. In fact, instead of sending out all the symbols of $A_2' \oplus B_1'$, $A_2' \oplus B_3'$ and $A_3' \oplus A_1'$, the server only needs to send out $\frac{6.5}{32}F$ symbols of each of them. In total, in the delivery phase, the server sends out $\frac{3}{32}F+\frac{6.5}{32}F\times3=\frac{22.5}{32}F$ symbols. The normalized rate is $\frac{45}{64}$, which is lower than $\frac{3}{4}$ obtained from \eqref{exact_un}.

We also try a different MDS code for this system. For a $(1.5F, F)$ MDS code, we calculate that the normalized traffic load as $\frac{25}{36}$, which is lower than the normalized traffic load obtained through $(2F, F)$ MDS code. We see in this example that the performance does not always have to improve with $r$. In this example, $r=1.5$ performs better than $r=2$.  \QEDA
\end{example}

We summarize the MDS coded prefetching based decentralized coded caching algorithm in Algorithm $1$. We use $[N]$ to represent $\{1,\dots,N\}$. Some remarks on Algorithm $1$ are as follows: In the placement phase, user $k$ caches $\frac{MF}{N}$ symbols of each MDS coded file. In total, $MF$ symbols are cached for each user; therefore, the memory size constraint is satisfied. Lines $11$ and $12$ determine a subset of the $K$ users called leaders as in \cite[Section IV]{yu2016exact}. Lines $19$ to $21$ show that if none of the leaders is related to the transmission, then the server does not need to send out this message, which is originally proposed in \cite[Section IV]{yu2016exact}. Line $22$ shows the coded multicasting message. Line $16$ accounts for at most how many symbols each user can get in iteration $|S|=j$. Line $17$ guarantees that each user does not get more symbols than they need. For example, in Example~\ref{ex3}, in iteration $|S|=2$, each user can get $\frac{18}{32}F$ symbols. However, only $\frac{13}{32}F$ symbols are needed to reconstruct the file each user requested. Lines $24$ to $26$ show that each user only require $F$ symbols to reconstruct the file they requested. For example, in Example~\ref{ex3}, the iteration stops at $|S|=2$.
\begin{algorithm}
	\caption{MDS assisted decentralized coded caching}\label{alg:MDS_dec}
	\begin{algorithmic}[1]
		\Procedure{Coded Prefetching} {$r$} \Comment{$(rF, F)$ MDS}
		\For {$n\in[N]$}
		\State encode file $W_n$ into MDS coded file, $W'_n$, with $rF$ coded symbols
		\EndFor
		\For {$k\in[K']$, $n \in [N]$}
		\State user $k$ uniformly and randomly chooses a subset of $\frac{MF}{N}$ symbols from each coded file, $W'_n$, to cache
		\State each user $k$ performs the prefetching independently
		\EndFor	
		\EndProcedure		
		\item[]
		\Procedure{Delivery} {$d_1, \dots, d_K$}
		\State $N(\mathbf{d})\gets$ distinct elements in $\mathbf{d}$
		\State $\mathcal{U}\gets$ users request different files, $|\mathcal{U}|=N(\mathbf{d})$
		\State $acc \gets \frac{MF}{N}$
		\For{$j=K,K-1,\dots,1$}
			\State $seg \gets |W'_{i,A}|$, for a $A \subset [K]: |A|=j-1$
			\State $acc_{new} \gets acc+ seg \times \binom{K-1}{j-1} $
			\State $incr \gets \frac{ \min\{seg \times \binom{K-1}{j-1} , F-acc\} }{ \binom{K-1}{j-1} }  $
			\For { $S\subset [K]: |S|=j$ }
				\If {$S \cap \mathcal{U} == \emptyset $ }
					\State continue
				\EndIf
				\State server sends $incr$ symbols of $\oplus_{k \in S} W'_{d_k,S \setminus \{k\}}$
			\EndFor
			\If{$acc_{new}>=F$}
				\State break
			\EndIf
			\State $acc \gets acc_{new}$
		\EndFor
		\EndProcedure		
	\end{algorithmic}
\end{algorithm}

Different from the coded prefetching schemes for centralized setting \cite{chen2014fundamental, tian2016caching, gomez2016fundamental}, the proposed scheme works for arbitrary $N$ and $K$. In fact, for decentralized setting, we cannot assume $N \leq K$, since the number of users making request in the delivery phase varies. By letting $r=1$ in line $1$ of Algorithm $1$, the proposed scheme reduces back to the original uncoded prefetching decentralized coded caching scheme in \cite{maddah2015decentralized}. The proposed algorithm can also be applied to different network topologies as that in \cite{maddah2015decentralized}.

We conclude this section with the following theorem.

\begin{theorem}
Consider the decentralized two-phase caching network with $N$ files in the server, each user with local memory size that can store $M$ files, and $K$ users making a request in the delivery phase. For large enough file size $F$, by applying a $(rF,F)$ MDS code and using Algorithm $1$, the following normalized rate is achievable,
\begin{align}
R_{\mathrm{dec}}=& \sum_{j=s+1}^{K} r q^{j-1} (1-q)^{K-j+1} \left(\binom{K}{j} - \binom{K-J}{j} \right) \notag \\
                 & +\frac{\left(1-A(s+1)\right)}{\binom{K-1}{s-1}} \left(\binom{K}{s} - \binom{K-J}{s} \right),   \label{exp_b}
\end{align}
where
\begin{align}
q=&\frac{M}{rN}, \quad J=\min\{N,K\},
\end{align}
and $s$ is defined through
\begin{align}
A(x)=\frac{M}{N}+\sum_{j=x}^{K}    r    q^{j-1}  (1-q)^{K-j+1}  \binom{K-1}{j-1},   \label{exp_a}
\end{align}
such that
\begin{align}
A(s+1) <  1, \quad A(s)   \geq  1   \label{exp_c}.
\end{align}
\end{theorem}
\begin{Proof}
$A(x)F$ is the accumulated number of symbols of the requested file by iteration $j=x$ starting from $j=K$. In the placement phase, each user stores $\frac{MF}{N}$ symbols of each file. Thus, we have the first term on the right hand side of \eqref{exp_a}. By law of large numbers, each encoded file segment is of size
\begin{align}
|W'_{n,A}| \approx rF q ^{|A|} (1-q) ^{K-|A|},
\end{align}
where $q=\frac{\frac{MF}{N}}{rF}=\frac{M}{rN}$.

In the delivery phase, we first focus on the iterations of $s+1 \leq j \leq K$. For the $jth$ iteration, the server transmits $\binom{K}{j} - \binom{K-J}{j}$ messages each with $rF q ^{j-1} (1-q) ^{K-j+1}$ symbols, which accounts for the first term on the right hand side of \eqref{exp_b}. The $kth$ user receives the broadcast message $\oplus_{k \in S} W'_{d_k,S \setminus \{k\}}$. Only the requested file segment, $W'_{d_k,S \setminus \{k\}}$, is unknown to this user; therefore, correct decoding can be performed. After each iteration, each user gets $r F   q^{j-1}  (1-q)^{K-j+1}  \binom{K-1}{j-1}$ symbols of the requested file; therefore, this accounts for the second term on the right hand side of \eqref{exp_a}.

In the last iteration, $j=s$, each user accumulates more than enough number of symbols to reconstruct the requested file. \eqref{exp_c} characterizes the last iteration. In the last iteration, the server only needs to transmits $F\frac{1-A(s+1)}{\binom{K-1}{s-1}}$ symbols of each broadcast message, since this is sufficient for each user to reconstruct the requested file. This gives the second term of the right hand side of \eqref{exp_b}.
\end{Proof}

\section{Numerical Results}

\subsection{Different Number of Users in the Delivery Phase}

\begin{figure}
	\centering
	\epsfig{file=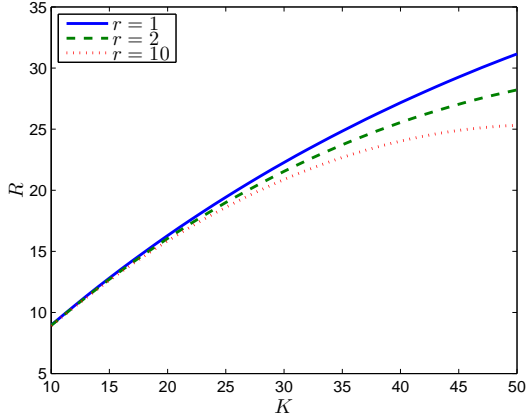,width=0.44\textwidth}
	\caption{Rate versus number of users $K$ in the delivery phase for $N=100$ and $M=2$.}
	\label{Fig_1}
\vspace*{-0.4cm}
\end{figure}

Consider the two-phase caching networks with $N=100$, $M=2$, and $K'=50$. To apply Algorithm $1$, we consider $(2F,F)$ and $(10F, F)$ MDS codes and compare their performance with the uncoded prefetching scheme. In the delivery phase, we consider that the number of users making request ranges from $10$ to $50$. Different from \cite{chen2014fundamental, tian2016caching, gomez2016fundamental}, this is the case of coded prefetching without the constraint of more users  than files. We show the numerical results in Fig.~\ref{Fig_1}, where we observe that longer MDS coded cases result in lower normalized rate. We also observe that the reduction in the transmission rate is more significant when the number of users making a request is larger.

\subsection{Rate Memory Trade-off}

\begin{figure}
	\centering
	\epsfig{file=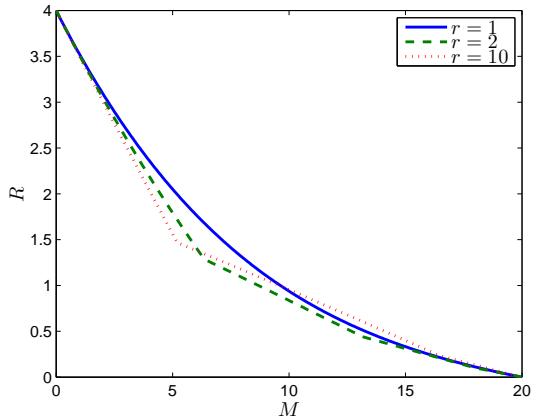,width=0.44\textwidth}
	\caption{Rate memory trade-off curve for $N=20$ and $K=4$.}
	\label{Fig_2}
\vspace*{-0.4cm}
\end{figure}

We consider the rate memory trade-off curve for the two-phase caching network with $N=20$ and $K=4$. We show the numerical results in Fig.~\ref{Fig_2}. Although $(2F,F)$ MDS coded case achieves lower normalized rate, $(10F,F)$ MDS coded case does not always achieve a lower rate. We observed a similar phenomenon in Example~\ref{ex3}, where $(1.5F,F)$ MDS coded case resulted in lower normalized rate than that of $(2F,F)$.

\begin{figure}
	\centering
	\epsfig{file=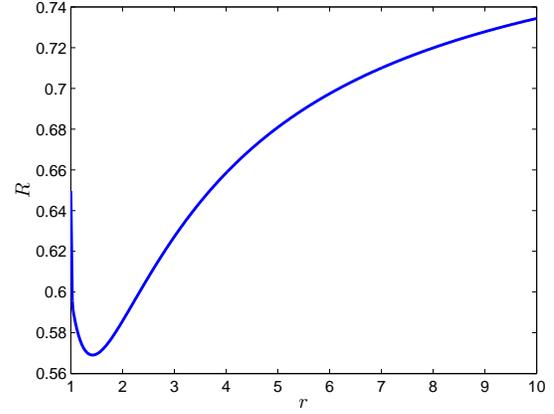,width=0.44\textwidth}
	\caption{Rate versus MDS code block length $r$ for $N=20$, $M=12$ and $K=4$.}
	\label{Fig_3}
\vspace*{-0.4cm}
\end{figure}

Since $N$ and $M$ are fixed for the two-phase caching networks, we can use an expected number of users $K$ in the delivery phase to choose a proper MDS code parameter, $r$, in advance. For example, following the same setting, for $N=20$ and $K=4$, consider the case $M=12$. We show in Fig.~\ref{Fig_3} that different coded MDS cases result in different normalized rates. We observe that longer MDS coded case does not necessarily result in lower normalized rate. From Fig.~\ref{Fig_3}, we conclude that we should choose $r \approx 1.5 $ in Algorithm $1$ for this case. Note that for $r=1$, Algorithm $1$ is the uncoded prefetching decentralized coded caching scheme in \cite{maddah2015decentralized}; therefore, Algorithm $1$ adds one more degree of freedom for us to lower the normalized rate.

\section{Conclusion}

We proposed a new decentralized coded caching scheme based on MDS coded prefetching, and obtained improved rate memory trade-off results. The proposed scheme is a generalization to the original uncoded prefetching decentralized coded caching scheme, and adds one more degree of freedom to the original scheme, which is the MDS code rate, to optimize.

\renewcommand{\baselinestretch}{0.985}

\end{document}